\documentclass [12pt, letterpaper]{article}
\usepackage{fullpage}
\usepackage{amsmath}
\usepackage{amssymb}
\usepackage{graphicx}
\usepackage{mathrsfs}
\usepackage{wrapfig}
\usepackage{boxedminipage}
\usepackage{epsfig}
\usepackage{setspace}
\usepackage{subfigure}
\usepackage{float}
\usepackage{hyperref}
\newcommand{\reef}[1]{(\ref{#1})}
\begin{document}
\begin{flushright}
\phantom{{\tt arXiv:}}
\end{flushright}

\bigskip
\bigskip
\bigskip

\begin{center} {\Large \bf Non--Unitary Fermionic }
  
  \bigskip

{\Large\bf  Quasinormal Modes at Zero Frequency}

\end{center}

\bigskip \bigskip \bigskip \bigskip

\centerline{\bf Tameem Albash}

\bigskip
\bigskip
\bigskip

  \centerline{\it Department of Physics and Astronomy }
\centerline{\it University of
Southern California}
\centerline{\it Los Angeles, CA 90089-0484, U.S.A.}

\bigskip

\centerline{\small \tt talbash[at]usc.edu}

\bigskip
\bigskip


\begin{abstract} 
\noindent 
We consider the dynamics of a probe fermion charged under a $U(1)$ Maxwell field and a two form potential $B_{(2)}$ in a five dimensional gravity background.  The gravity background is constructed from a new solution we find of type IIB supergravity.  This new solution is expected to be dual to non--commutative Yang--Mills theory in the 't Hooft limit with global $U(1)$ currents.  We study the zero frequency, near horizon behavior of the fermion, where the equations of motion reduce to that of two interacting fermions in AdS$_2$ with an electric field.  We show that the operator dimensions in the AdS$_2$ space are complex, leading to the two components of the retarded Green's function in the dual theory to be complex conjugates of each other.  In order to preserve unitarity, this result implies there are no zero frequency quasinormal modes in our system.  This has important implications for generalizations of recent holographic Fermi liquid setups with AdS$_2$ regions, as it suggests that infinite lifetime excitations can have energies above/below the chemical potential.  Therefore, the Fermi energy may not be uniquely set by the chemical potential.  Furthermore, since the gravity background breaks rotational symmetry along the spatial directions of the dual Yang--Mills theory, we do not expect the Fermi surface to be spherical in shape in momentum space.
\end{abstract}
\newpage \baselineskip=18pt \setcounter{footnote}{0}

\section{Introduction}
%
The formulation of the AdS/CFT correspondence \cite{Maldacena:1997re, Witten:1998qj,Gubser:1998bc} has provided the best understood example of holography.  Among the novel features of this duality is that it is an example of a strong/weak coupling duality, mapping strongly coupled physics in a non--gravitational theory to weakly coupled physics in a gravitational theory and \emph{vice versa}.  The correspondence has been generalized to include various parameters of interest in physics, such as temperature \cite{Witten:1998zw} and finite $U(1)$ chemical potential \cite{Chamblin:1999tk}, which are dual to having black holes with event horizons and electrical charge respectively.  Such generalizations of the correspondence (referred generally as gauge/gravity dualities) provide us with a new tool to probe strongly coupled physics.  In particular, various novel phenomena in condensed matter physics are believed to be strongly coupled (see for example, refs.~\cite{varma-2002-361, PhysRevB.78.035103}), and the correspondence may be able to shed some light on these phenomena.

To that end, there has been a lot of interest in building gravity models that capture some of the physics of interest in condensed matter systems.  An excellent example of this is the work done on holographic superconductivity \cite{Gubser:2008px,Hartnoll:2008vx}, where an abelian scalar--Maxwell sector in the gravity theory can be shown to spontaneously break the global $U(1)$ symmetry in the dual gauge theory.  Although the broken symmetry is global in nature (so technically it is a superfluid), it can be weakly gauged and shown to still capture the relevant physics of superconductivity \cite{Hartnoll:2008kx}.

Another example is recent work on studying probe fermions in Reissner--Nordstr\"om backgrounds \cite{Lee:2008xf,Liu:2009dm}.    At zero temperature, studies reveal the presence of a pole in the spectral function reminiscent of quasiparticle excitations near Fermi surfaces.  The pole is found at $\omega = 0$ and finite spatial momentum, where $\omega$ is identified to be the energy above/below the Fermi energy $E_F$, which is set by the chemical potential $\mu$.  The pole exhibits a dispersion relation that deviates drastically from that expected from Landau theory, which fits nicely with the fact that it must be strongly--coupled in nature by the gauge/gravity correspondence.  In ref.~\cite{Faulkner:2009wj}, it was shown that many of the properties of the quasiparticle pole are determined by the AdS$_2$ near horizon, $\omega \to 0$ limit of the geometry.  Furthermore, the peak in the spectral function appears to show the expected broadening behavior for a Fermi surface excitation at finite temperature \cite{Cubrovic:2009ye}.

Of course, these results come with several caveats.  It is known that the Reissner--Nordstr\"om background is not the thermodynamically favored state at zero temperature \cite{Chamblin:1999tk}; in addition, recent holographic superconductor studies have shown that a phase transition occurs before $T = 0$ \cite{Gubser:2008px,Hartnoll:2008vx,Hartnoll:2008kx}, so even at finite temperature one would perhaps expect a ``hairy'' black hole and not Reissner--Nordstr\"om.  For these reasons and others, either the Reissner--Nordstr\"om background is not the correct background to be studying or the Fermi surface may simply not be accessible in these systems.

Nevertheless, the probe fermion in Reissner--Nordstr\"om provides us with a simple system that captures many of the features of a Fermi liquid that we hope to extract.  By studying what gives rise to these desired features provides us with valuable insight into searching for realistic gravity duals of Fermi liquids (see for example, ref.~\cite{Hartnoll:2009ns}).  The phenomenological approach to the holographic superconductor is a good example of this process where the consistent embedding into string theory \cite{Gubser:2009qm,Gauntlett:2009dn,Gubser:2009gp} was only found after the phenomenological studies.  It is with this philosophy that we approach our problem.

In further studies of the holographic Fermi liquid setup of ref.~\cite{Liu:2009dm}, it was shown in ref.~\cite{Albash:2009wz,Albash:2010yr} that in the presence of a background magnetic field, the quasiparticle pole is shifted away from $\omega = 0$ for a certain class of solutions.  This deviation from $\omega = 0$ was observed by solving the partial differential equations for the fermions numerically, so we wish to study a system with analogous properties but with simpler equations of motion that we can study analytically.  To that end, we propose to accomplish this with a background with a two form potential $B_{(2)}$ with components along the AdS$_5$ space.  We want our background to have an AdS$_2$ near horizon geometry in order to relate our observations to the holographic Fermi liquid results of ref.~\cite{Faulkner:2009wj}.  This can be accomplished by having an electrically charged black hole.  We find a new ten dimensional solution to the type IIB supergravity action with these properties; the background can be generated by performing the T--duality transformations of refs.~\cite{Maldacena:1999mh,Hashimoto:1999ut} on the spinning D3--brane solution \cite{Kraus:1998hv,Cvetic:1999xp}.  This background is expected to be dual to non--commutative Yang--Mills theory in the 't Hooft limit with global $U(1)$ currents turned on.  From the ten dimensional solution, we construct a five dimensional background that inherits the features we desire.

We probe the five dimensional background with a Dirac field charged under the $U(1)$ gauge field and the $B_{(2)}$ field.  In the near horizon limit, the four component Dirac field can be viewed as a pair of two component Dirac fields in AdS$_2$ with an interaction term controlled by $B_{(2)}$.  This interaction term leads to some novel results for the quasinormal modes of the system with implications to the excitations of the underlying Fermi surface, which we summarize next.  The interaction splits the degeneracy in the operator dimension that exists when the $B_{(2)}$ term is absent.  The operator dimension is found to be complex, rather than simply pure real or pure imaginary.  Assuming the retarded Green's function retains the form proposed in ref.~\cite{Faulkner:2009wj} and assuming one can extract the value of the sources and vacuum expectations values (vevs) of the dual operators, we find that the two components of the retarded Green's function matrix at $\omega = 0$ are complex conjugates of each other.  This result implies that we generically have a negative spectral function (proportional to the imaginary part of the retarded Green's function) at $\omega = 0$, which is disallowed by unitarity.  Since the energy for infinite lifetime excitations is given by $E = \omega  + \mu$, this suggests that that the Fermi energy is not equal to the chemical potential.  Finally, since the probe fermion is in a background that does not have rotational symmetry in the three spatial directions perpendicular to the AdS radial coordinate and the time coordinate, we expect the Fermi surface not to be spherical in shape in momentum space.

The paper is organized as follows.  In section \ref{sec:geometry}, we present our ten dimensional supergravity solution and find a five dimensional action whose solution captures the elements of the ten dimensional solution that we desire.  In section \ref{sec:probe}, we present our probe fermion calculation and work out the equations of motion for the fermion.  In section \ref{sec:w=0}, we study the $\omega \to 0$ limit of our probe fermion and relate it to the physics of spinors in AdS$_2$.  In section \ref{sec:speculate}, we make our assumption on the form of the retarded Green's function and study the consequences that result by using properties of the solutions we have found.  Finally, in section \ref{sec:conclusions} we summarize our findings and conclusions.
%
\section{The Gravity Background} \label{sec:geometry}
%
\subsection{The Ten Dimensional Solution}
The type IIB action in ten dimensions is given by:
\begin{eqnarray}
S_{IIB} &=& \frac{1}{2 \kappa_0^2} \int d^{10} x \sqrt{-G} \left\{ e^{-2 \Phi} \left( \mathcal{R} +4 \partial_\mu \Phi \partial^\mu \Phi - \frac{1}{2} |H_{(3)}|^2 \right) - \frac{1}{2} |\tilde{F}_{(3)}|^2 - \frac{1}{4} |\tilde{F}_{(5)}|^2 \right\} \nonumber \\
&& - \frac{1}{4 \kappa_0^2} \int C_{(4)} \wedge H_{(3)} \wedge F_{(3)} \ ,
\end{eqnarray}
where we are using the convention that:
\begin{equation*}
|G_{(p)}|^2 = \frac{1}{p!} \left(G_{(p)}\right)^{\mu_1  \dots  \mu_p} \left(G_{(p)}\right)_{\mu_1  \dots  \mu_p} \ ,
\end{equation*}
\begin{equation*}
F_{(p)} = d C_{(p-1)} \ , \quad H_{(3)} = d B_{(2)} \ , \quad \tilde{F}_{(3)} = F_{(3)} - C_{(0)} \wedge H_{(3)} \ , \quad \tilde{F}_{(5)} = F_{(5)} + B_{(2)} \wedge F_{(3)} \ .
\end{equation*}
We propose the following new ten dimension solution (in string frame):
\begin{eqnarray} 
\frac{ds^2}{R^2} &=&-v^2  f(v) dt^2 + v^2 d x_1^2 + \frac{v^2}{1+ a^4 v^4} \left( dx_2^2 + dx_3^2 \right)+ \frac{dv^2}{v^2  f(v) } \nonumber \\
&& + \sum_{i=1}^{3} \left( d \mu_i^2 + \mu_i^2 \left( d \phi_i -  A_t dt \right)^2 \right) \ , \nonumber \\
f(v) &=& 1 - \frac{v_H^4(v^2 - q^2)}{v^6} \ , \quad C_{(2)} = - \frac{R^2}{g_s} a^2 \left( v^4 d t + q v_H^2 \sum_{i=1}^3 \mu_i^2 d \phi_i \right) \wedge d x_1 \ , \nonumber \\
B_{(2)} &=& R^2 \frac{a^2 v^4}{1+ a^4 v^4} d x_2 \wedge d x_3 \ , \quad e^{2 \Phi} =  \frac{g_s^2}{1+ a^4 v^4}\ , \quad A_t = \frac{q v_H^2}{ v^2} \ , \nonumber \\
\mu_1 &=& \sin \theta \ , \quad \mu_2 = \cos \theta \sin \psi \ , \quad \mu_3 = \cos \theta \cos \psi \ , \quad  0 \leq \theta \leq \pi/2 \ , \quad 0 \leq \psi \leq 2 \pi \ , \nonumber \\
\label{eqt:10d_background}
C^{(4)}_{ t x_1 x_2 x_3} &=& \frac{R^4}{g_s} v^4 \ , \quad
C^{(4)}_{\psi \phi_2 \phi_3 \phi_1} = \frac{R^4}{g_s} \mu_2 \mu_3 \cos^2 \theta \ , \quad C^{(4)}_{\phi_i x_1 x_2 x_3} = \frac{R^4}{g_s} v_H^2 q \mu_i^2 \ , \quad i=1,2,3 \ , \nonumber \\
C^{(4)}_{t \psi \phi_1 \phi_3} &=& C^{(4)}_{t \psi \phi_2 \phi_1} = \frac{R^4}{g_s} A_t \mu_1^2 \mu_2 \mu_3  \ , \quad C^{(4)}_{t \theta \phi_1 \phi_3} = \frac{R^4}{g_s} A_t \mu_1 \mu_3 \cos \psi \ , \nonumber \\
 C^{(4)}_{t \theta \phi_1 \phi_2} &=& \frac{R^4}{g_s} A_t \mu_1 \mu_2 \sin \psi  \ , \quad C^{(4)}_{t \psi \phi_2 \phi_3} = \frac{R^4}{g_s} A_t  \mu_2 \mu_3 \cos^2 \theta   \ . 
\end{eqnarray}
In the coordinates we are using, $(v,q,v_H)$ have units of inverse length and $(a, t, x_i)$ have units of length.  As indicated in the introduction, we can construct this solution by following the sequence of T--duality transformations proposed in refs.~\cite{Maldacena:1999mh,Hashimoto:1999ut}.  By starting with the spinning D3--brane solution with equal angular momenta along the Cartan $U(1)^3$ of $SO(6)$ \cite{Kraus:1998hv,Cvetic:1999xp}, the solution is constructed by performing a T--duality transformation along the $x_2$ direction, which results in smeared D2--branes on a tilted torus, then T-dualizing back on the $x_3$ direction.
Our ten dimensional solution in equation \reef{eqt:10d_background} limits nicely to known solutions.  By setting $q = 0$, we recover the non--extremal solution of ref.~\cite{Maldacena:1999mh}\footnote{We use a slightly different convention than ref.~\cite{Maldacena:1999mh} such that our solution for $C_{(2)}$ has a negative sign.}, which is dual to strongly coupled non--commutative Yang--Mills theory in the 't Hooft limit.  The non--commutativity in the dual field theory is in the $(x_2, x_3)$ directions, \emph{i.e.}:
\begin{equation}
\left[ x_2 , x_3 \right] \sim a^2 \ . 
\end{equation}
By setting $a = 0$, we recover the spinning D3--brane solution with equal angular momenta, which is dual to strongly coupled Yang--Mills theory in the 't Hooft limit with global $U(1)$ currents \cite{Chamblin:1999tk}.  Therefore, we expect our background to be dual to strongly coupled non--commutative Yang--Mills theory in the 't Hooft limit with $U(1)$ global currents.
%
\subsection{Constructing the Five Dimensional Solution and Action}
%
The ten dimensional solution (in string frame) in equation \reef{eqt:10d_background} has the same off--diagonal components in the metric as the spinning D3--brane with equal momenta.  For this reason, it is natural to expect that the dimensional reduction on the five sphere should give the same result as in the spinning D3--brane case, which is to produce a non--zero time component for the gauge field in the five dimensional theory.  The $B_{(2)}$ form in the ten dimensional solution has no support on the five sphere, so we expect it to be untouched by the reduction.  The $C_{(2)}$ form \emph{does} have support on the five sphere, so we can expect that the reduction introduces a new vector field in the five dimensional theory.  Finally, since this reduction was on the string frame solution, we need to redefine the metric to be in Einstein frame.  This is done by:
\begin{equation} \label{eqt:metric_st_ei}
G_{\mu \nu} = e^{- \frac{4}{3} \Phi} G_{\mu \nu}^{st} \ .
\end{equation}
On these grounds, we expect the five dimensional solution (in Einstein frame) to be given by:
\begin{eqnarray} \label{eqt:5d_background}
\frac{ds^2}{R^2}  &=& g_s^{-4/3} \left(1+a^4 v^4 \right)^{2/3} \left(-v^2  f(v) dt^2 + v^2 d x_1^2 + \frac{v^2}{1+ a^4 v^4} \left( dx_2^2 + dx_3^2 \right)+ \frac{dv^2}{v^2  f(v) }   \right) \ , \nonumber \\
C_{(2)} &=& - \frac{R^2 a^2  v^4}{g_s} d t  \wedge d x_1 \ , \quad B_{(2)} = R^2 \frac{a^2 v^4}{1+ a^4 v^4} d x_2 \wedge dx_3  \ , \nonumber \\
e^{2 \Phi} &=& g_s^2 e^{2 \tilde{\Phi} } =  \frac{g_s^2}{1+ a^4 v^4} \ , \quad A_{(1)} = \frac{R q v_H^2}{ v^2} dt   \ , \quad F_{(2)} = d A_{(1)} =  \frac{2 R q v_H^2}{ v^3} dt \wedge dv \ , \nonumber \\
\tilde{A}_{(1)}  &=& \frac{2 R a^2 v_H^2 q}{g_s} d x_1 \ , \quad  \tilde{F}_{(2)} = d \tilde{A}_{(1)} = 0 \ , \quad f(v) = 1 - \frac{v^2 - q^2}{v^6} v_H^4 \ ,
\end{eqnarray}
where for now, we explicitly keep powers of $g_s$.  We wish to construct the five dimensional action whose equations of motion give rise to the solution in equation \reef{eqt:5d_background}.  We know that the spinning D3--brane solution with equal momenta reduces to a solution of Einstein--Maxwell theory with a cosmological constant, so we expect our solution to be a heavily decorated version of this.  We motivate the various pieces of it in what follows.  First, we can expect the kinetic terms for the dilaton, $B_{(2)}$, and $C_{(2)}$ fields to be largely unchanged from what they were in the ten dimensional action, except for their overall coefficient that is modified by the metric redefinition of equation \reef{eqt:metric_st_ei}.  We write this piece as:
\begin{equation} \label{eqt:S1}
S_1 \sim   \int d^5 x \sqrt{-G} \left( \mathcal{R} - \frac{4}{3} \partial_\mu \tilde{\Phi} \partial^{\mu} \tilde{\Phi} - \frac{e^{-\frac{8}{3}\tilde{\Phi}}}{12} H^{(3)}_{\mu \nu \lambda} H_{(3)}^{\mu \nu \lambda} -\frac{e^{-\frac{2}{3}\tilde{\Phi}}}{12} F^{(3)}_{\mu \nu \lambda} F^{\mu \nu \lambda}_{(3)}  \right) \ .
\end{equation}
In five dimensions, we can write Chern--Simons terms involving $B_{(2)}$, $C_{(2)}$, $A_{(1)}$, $\tilde{A}_{(1)}$, and their field strengths.  Therefore, we can write another piece of the action as:
\begin{eqnarray} \label{eqt:S2}
S_2 &\sim&  \frac{2}{g_s R} \int \left( B_{(2)} \wedge F_{(3)} - C_{(2)} \wedge H_{(3)} \right)  - \frac{1}{R} \int d  \left( \frac{1}{g_s} B_{(2)} - C_{(2)} \right) \wedge A_{(1)} \wedge \tilde{A}_{(1)}  \nonumber \\
&& + \frac{1}{R} \int \left( \left( \frac{1}{g_s} B_{(2)} - C_{(2)} \right) \wedge F_{(2)} \wedge \tilde{A}_{(1)} +  \left( \frac{1}{g_s} B_{(2)} - C_{(2)} \right) \wedge A_{(1)} \wedge \tilde{F}_{(2)} \right) \ ,
\end{eqnarray}
where the particular form is motivated by the $sl(2,\mathbb{R})$ structure present in ten dimensions as well as gauge invariance.  Finally, we expect kinetic terms and possible interaction terms for the fields $A_{(1)}$ and $\tilde{A}_{(1)}$.  We write these as:
\begin{eqnarray} \label{eqt:S3}
S_3 &\sim& \int d^5 x \sqrt{-G} \left(   - \frac{1}{4} \left( e^{-\frac{4}{3} \Phi} + \frac{2}{g_s^2} e^{ \frac{2}{3} \Phi} \right) F^{(2)}_{\mu \nu} F_{(2)}^{\mu \nu} - c \tilde{F}^{(2)}_{\mu \nu} \tilde{F}^{\mu \nu}_{(2)} - \frac{1}{R^2} e^{2 \Phi} \tilde{A}^{(1)}_\mu \tilde{A}_{(1)}^{\mu}  \right. \nonumber \\
&& \left. + \frac{20}{R^2}e^{\frac{4}{3} \Phi} - \frac{8}{g_s^2 R^2 } e^{\frac{10}{3} \Phi} \right) \ .
\end{eqnarray}
The exact form of the coefficients in equation \reef{eqt:S3} are chosen so that equation \reef{eqt:5d_background} is a solution of the equations of motion derived from the action $S_1 + S_2 + S_3$.  Because of this \emph{ad hoc} construction,  we are unable to determine the overall factor $c$ of the kinetic term of $\tilde{A}_{(1)}$ as it does not contribute to any of the equations of motion for our solution.  We wish to emphasize that all the action terms have an overall power of $g_s^{-2}$, which is exactly what we should expect in order to define Newton's constant $G_N$ in five dimensions.  Therefore, we can make the following redefinition:
\begin{equation}
G_{\mu \nu} \to g_s^{-4/3} G_{\mu \nu} \ , \quad C_{(2)} \to \frac{1}{g_s} C_{(2)}  \ , \quad \tilde{A}_{(1)} \to \frac{1}{g_s} \tilde{A}_{(1)} \ , 
\end{equation}
such that our five dimensional action is given by:
\begin{eqnarray}
16 \pi G_5 S_{5d} &=& \int d^5 x \sqrt{-G} \left( \mathcal{R} - \frac{4}{3} \partial_\mu \tilde{\Phi} \partial^{\mu} \tilde{\Phi} + \frac{20}{R^2}e^{\frac{4}{3} \tilde{\Phi}} - \frac{8}{R^2 } e^{\frac{10}{3} \tilde{\Phi}} - \frac{e^{-\frac{8}{3}\tilde{\Phi}}}{12} H^{(3)}_{\mu \nu \lambda} H_{(3)}^{\mu \nu \lambda}  \right. \nonumber \nonumber \\
&& \left.-\frac{e^{-\frac{2}{3}\tilde{\Phi}}}{12} F^{(3)}_{\mu \nu \lambda} F^{\mu \nu \lambda}_{(3)}  - \frac{1}{4} \left( e^{-\frac{4}{3} \tilde{\Phi}} + 2 e^{ \frac{2}{3} \tilde{\Phi}} \right) F^{(2)}_{\mu \nu} F_{(2)}^{\mu \nu} - c \tilde{F}^{(2)}_{\mu \nu} \tilde{F}^{\mu \nu}_{(2)} - \frac{1}{R^2} e^{2 \tilde{\Phi}} \tilde{A}^{(1)}_\mu \tilde{A}_{(1)}^{\mu}  \right) \nonumber \\
&&+ \frac{2}{R} \int \left( B_{(2)} \wedge F_{(3)} - C_{(2)} \wedge H_{(3)} \right)  - \frac{1}{R} \int d  \left( B_{(2)} - C_{(2)} \right) \wedge A_{(1)} \wedge \tilde{A}_{(1)}  \nonumber \\
&& + \frac{1}{R} \int \left( \left( B_{(2)} - C_{(2)} \right) \wedge F_{(2)} \wedge \tilde{A}_{(1)} +  \left( B_{(2)} - C_{(2)} \right) \wedge A_{(1)} \wedge \tilde{F}_{(2)} \right) \nonumber \ .
\end{eqnarray}
where we have used that $G_5 = \frac{\pi}{2 R^3 N^2} = (16 \pi R^5 \mathrm{Vol}(S^5) / (2 \kappa_0^2 g_s^2))^{-1}$.  A solution to the equations of motion derived from this action is equation \reef{eqt:5d_background} with $g_s$ set to one.
Before proceeding with our probe computation, it is convenient to make the following field redefinition for the $U(1)$ gauge field $A_{(1)}$ such that:
\begin{equation}
A_{(1)} \to \frac{R}{\sqrt{3}} A_{(1)} \ .
\end{equation}
Furthermore, it is convenient to work in terms of dimensionless coordinates and parameters.  To this end, let us define:
\begin{equation}
z = \frac{v_H}{v} \ , \quad q \to v_H q \ , \quad a \to v_H^{-1} a \ , \quad (t, x_i) \to  (\frac{t}{v_H}, \frac{x_i}{v_H}) \ .
\end{equation}
Under this change, the background can be written as:
\begin{equation*}
\frac{ds^2}{R^2} = \left( 1 + \frac{a^4}{z^4} \right)^{2/3} \left( -\frac{f(z)}{z^2}dt^2 + \frac{dx_1^2}{z^2} + \frac{1}{1 + \frac{a^4}{z^4}} \frac{1}{z^2} \left( dx_2^2 + dx_3^2 \right) + \frac{dz^2}{z^2 f(z)} \right) \ ,
\end{equation*}
\begin{equation} \label{eqt:background_z}
A^{(1)}_t= \sqrt{3} q \left( z^2 - z_{0}^2 \right) \ , \quad B^{(2)}_{x_2 x_3} = \frac{a^2 R^2}{a^4 + z^4} \ ,  \quad C^{(2)}_{t x_1} = - R^2 a^2 \left( \frac{1}{z^4} - \frac{1}{z_0^4} \right) \ , \quad e^{2 \tilde{\Phi} } =  \frac{1}{1+ \frac{a^4}{ z^4}} \ ,
\end{equation}
where we have included a constant gauge shift for $A_t$ and $C_{t x_1}$ in terms of the position of the event horizon $z_0$ such that $A_t(z_{0}) = 0$ and $C^{(2)}_{t x_1}(z_0) = 0$ to ensure that the norm of these fields is regular at the event horizon \cite{Kobayashi:2006sb}.  In these coordinates, $f(z)$ is given by:
\begin{equation}
f(z) = q^2 z^6 -z^4 + 1 \ .
\end{equation}
The Hawking temperature is given by the usual Gibbons--Hawking calculus \cite{Gibbons:1979xm}:
\begin{equation}
T = \frac{z_0^3}{2 \pi} \left( 3 q^2 z_0^2 - 2 \right) \ .
\end{equation}
For our work, we are interested in working at zero temperature in order to have an AdS$_2$ near horizon geometry.  This corresponds to taking the extremal limit of our background, which requires, in our dimensionless parameters:
\begin{equation}
q = \left( \frac{4}{27} \right)^{1/4} \ .
\end{equation}
With this choice, the event horizon is at a radius of $z_0 = 3^{1/4}$, and $f(z)^{-1}$ exhibits the expected double pole at the event horizon for the extremal limit:
\begin{equation} \label{eqt:f_zero}
f(z) = q^2 \left( z - z_0 \right)^2 \left( z + z_0 \right)^2 \left( z^2 + \frac{z_0^2}{2} \right) \ .
\end{equation}
The near horizon limit is taken as follows:
\begin{equation} \label{eqt:limit}
t = \frac{\tau}{\lambda}  \ , \quad z_0 -z = \frac{\sqrt{3} \lambda }{12 \xi} \ , \quad z \to z_0 \ , \quad  \lambda \to 0, \quad \xi \ \mathrm{finite} \ ,
\end{equation}
where $\lambda$ is some parameter.  This yields:
\begin{equation}
f(z) \to \frac{\sqrt{3} \lambda^2}{12 \xi^2} \ , \quad A^{(1)}_t \to - \frac{\lambda}{\sqrt{6} \xi} \equiv \lambda \frac{e_d}{\xi} \ , \quad  C^{(2)}_{t x_1} \to - \frac{R^2 a^2 z_0 \lambda }{9 \xi} \ .
\end{equation}
In this limit, our background becomes:
\begin{eqnarray}
ds^2  &=&  \left( 1 + \frac{a^4}{z_0^4} \right)^{2/3} \left( \frac{R^2}{12 \xi^2} \left( - d \tau^2 + d \xi^2 \right) + \frac{R^2}{z_0^2} d x_1^2 + \frac{1}{1+ \frac{a^4}{z_0^4}} \frac{R^2}{z_0^2} \left( dx_2^2 + dx_3^2 \right) \right) \nonumber \\
B^{(2)}_{x_2 x_3} &=& \frac{a^2 R^2}{a^4 + z_0^4} \ , \quad C^{(2)}_{\tau x_1} = - \frac{R^2 a^2 z_0}{9 \xi} \ , \quad A^{(1)}_{\tau} = \frac{e_d}{\xi} \ ,  \quad e^{2 \tilde{\Phi} } =  \frac{1}{1+ \frac{a^4}{ z_0^4}} \ .
\end{eqnarray}
By defining:
\begin{equation}
R_2^2 = \frac{R^2}{12} \left( 1 + \frac{a^4}{z_0^4} \right)^{2/3} \ , \quad (x_2, x_3) \to \left( \frac{x_2}{1+ \frac{a^4}{z_0^4}} ,   \frac{x_3}{1+ \frac{a^4}{z_0^4}} \right) \ ,
\end{equation}
we find that the near horizon geometry of our background is AdS$_2  \times \mathbb{R}^3$ with flux, where the AdS$_2$ has radius $R_2$.  We emphasize that the presence of the flux continues to break the $SO(3)$ symmetry in the three spatial directions $x_{1,2,3}$.
%
\section{The Probe Fermion} \label{sec:probe}
%
Let us now consider the Dirac action for a fermion in the zero temperature background given in equations \reef{eqt:background_z} and \reef{eqt:f_zero}, and let us assume that this fermion is coupled to the $U(1)$ gauge field and the $B_{(2)}$ field as follows:
\begin{equation}
S_D \sim i \int d^5 x  \left( \bar{\Psi} \Gamma^{\mu} \mathcal{D}_\mu \Psi + \frac{1}{8R} e^{-\frac{4}{3} \tilde{\Phi}}  \bar{\Psi} B^{(2)}_{\mu \nu} \Gamma^{\mu \nu}  \Psi- m \bar{\Psi} \Psi \right) \ ,
\end{equation}
where $\mathcal{D}_\mu$ and $\Gamma^{\mu \nu}$ are given by:
\begin{equation}
\mathcal{D}_\mu = \partial_\mu +\frac{1}{4} \omega_{\mu \underline{a} \underline{b}} \Gamma^{\underline{a} \underline{b}} - i g A^{(1)}_\mu   \ , \quad \Gamma^{\mu \nu} = \Gamma^{[\mu}\Gamma^{ \nu]} = \frac{1}{2} \left[ \Gamma^{\mu}, \Gamma^{\nu} \right] \ .
\end{equation}
We use the convention where underlined indices are in the tangent space and indices that are not underlined are in the bulk spacetime.  Our choice for the form of the coupling term between $B_{(2)}$ and the fermionic field is motivated by the five dimensional $\mathcal{N}=8$ supergravity action \cite{Gunaydin:1984qu,Gunaydin:1985cu,Pernici:1985ju}, where the fermions couple to the potentials and not the field strengths .  The classical equation of motion for the field $\Psi$ is given by:
\begin{equation}
\left( \Gamma^{\mu} \mathcal{D}_\mu + \frac{1}{8R}e^{-\frac{4}{3} \tilde{\Phi}} B^{(2)}_{\mu \nu} \Gamma^{\mu \nu}-  m \right) \Psi = 0 \ .
\end{equation}
Because the non--commutativity prevents us from setting the momenta in the $(x_2,x_3)$ to zero, we consider an ansatz where the momentum in the $x_1$ direction vanishes\footnote{Our ansatz would study only the $k_1 = 0$ slice of the underlying Fermi surface (if it exists).}:
\begin{equation}
\Psi =  \left( -G^{zz} \det G  \right)^{-1/4} e^{-i \omega t  + i k_2 x_2 + i k_3 x_3} \left( \begin{array}{c} \phi_1 \\ \phi_2 \end{array} \right) \ ,
\end{equation}
 and we choose the following basis for our gamma matrices:
\begin{equation}
\Gamma^{\underline{t}} = \left( \begin{array}{cc}
 i \sigma_x & 0 \\
0 &  i \sigma_x
\end{array} \right) \ , \quad
\Gamma^{\underline{{x_1}}}= \left( \begin{array}{cc}
0 & - \sigma_y \\
-\sigma_y &  0
\end{array} \right) \ , \quad
\Gamma^{\underline{{x_2}}} = \frac{1}{\sqrt{k_2^2 + k_3^2}} \left( \begin{array}{cc}
- k_2 \sigma_y &- i k_3 \sigma_y  \\
 i k_3  \sigma_y & k_2 \sigma_y
\end{array} \right) \ , \nonumber
\end{equation}
\begin{equation}
\Gamma^{\underline{{x_3}}}= \frac{1}{\sqrt{k_2^2 + k_3^2}} \left( \begin{array}{cc}
- k_3 \sigma_y & i k_2 \sigma_y  \\
 - i k_2  \sigma_y & k_3 \sigma_y 
 \end{array} \right) \ , \quad
\Gamma^{\underline{z}}= \left( \begin{array}{cc}
-  \sigma_3 & 0 \\
0 & -\sigma_3
\end{array} \right) \ , \nonumber
\end{equation}
where $\sigma_{x,y,z}$ are the Pauli matrices.  This choice has the benefit of simplifying the following term:
\begin{equation}
i k_2 \Gamma^{\underline{x}_2} + i k_3 \Gamma^{\underline{x}_3} = \sqrt{ k_2^2 + k_3^2}  \left( \begin{array}{cc}
-  i \sigma_y & 0 \\
0 & i \sigma_y
\end{array} \right)   \ .
\end{equation}
The equations of motion in this basis give:
\begin{equation*}
\left( \sqrt{f(z)} \partial_z + \frac{m R}{z}\left( 1 + \frac{a^4}{z^4} \right)^{1/3}  \left( \begin{array}{cc} \sigma_z & 0 \\ 0 & \sigma_z \end{array} \right)   -  u \left( \begin{array}{cc} i \sigma_y & 0 \\ 0 & i \sigma_y \end{array} \right) + k \sqrt{1+ \frac{a^4}{z^4}} \left( \begin{array}{cc}  \sigma_x & 0 \\ 0 & - \sigma_x \end{array} \right) \right.
\end{equation*}
\begin{equation} \label{eqt:eom_outer}
\left.+  \frac{ i a^2 }{4 z^3} \left( \begin{array}{cc} 0 & \sigma_z \\ \sigma_z & 0 \end{array} \right)   \right) \left( \begin{array}{c} \phi_1 \\ \phi_2 \end{array} \right) = 0 \ , 
\end{equation}
where we have denoted $ k \equiv \sqrt{ k_2^2 + k_3^2}$ and $u = \frac{1}{\sqrt{f(z)}} \left( \omega + g A_t \right)$ for simplicity. The last term of equation \reef{eqt:eom_outer} is the contribution from the $B_{(2)}$ potential.  This term appears as an interaction term between $\phi_1$ and $\phi_2$ in this basis.  In addition, this term is the only imaginary term in the equations of motion. If we complex conjugate the equations of motion, we recover the original equations if we define:
\begin{equation} \label{eqt:cc}
\left( \begin{array}{c} \phi_1^\ast \\ \phi_2^\ast \end{array} \right) =  \left( \begin{array}{cc}  -1_{2 \times 2} & 0 \\ 0 & 1_{2 \times 2} \end{array} \right) \left( \begin{array}{c} \phi_1 \\ \phi_2 \end{array} \right) \ .
\end{equation}
%
\section{The Near Horizon Limit} \label{sec:w=0}
%
Let us take the near horizon limit given by equation \reef{eqt:limit} but with $\lambda$ taken to be equal to $\omega$.  We  define:
\begin{equation} \label{eqt:def}
\frac{z_0}{\sqrt{12}} \sqrt{1+ \frac{a^4}{z_0^4}} k \equiv R_2 \bar{m} \ , \quad \frac{a^2}{24} \equiv R_2 m_B \ ,
\end{equation}
such that the equations of motion for the fermion fields $\phi_{1,2}$ in the near horizon limit can be written as:
\begin{equation} \label{eqt:AdS2_v1}
  \xi \partial_\xi \phi_{^1_2} + m R_2  \sigma_z \phi_{^1_2} - i \xi \left( 1 + \frac{g e_d}{ \xi} \right) \sigma_y \phi_{^1_2} \pm  R_2 \bar{m} \sigma_x \phi_{^1_2}  + i R_2 m_B \sigma_z \phi_{^2_1}= 0 \ , 
\end{equation}
where the upper and lower subscripts on the index $\phi$ are associated with the upper and lower sign respectively.  We can recover the same equations by considering a pair of fermions in AdS$_2$.  This can be accomplished with the following action:
\begin{eqnarray} \label{eqt:action_AdS2}
S &=& i \int d^2 x \sqrt{-g} \left[ \bar{\psi}_1 \Gamma^{\alpha} \mathcal{D}_\alpha \psi_1 + \bar{\psi}_2 \Gamma^{\alpha} \mathcal{D}_\alpha \psi_2 - m \left( \bar{\psi}_1 \psi_1 + \bar{\psi}_2 \psi_2 \right) \right. \nonumber\\
&& \left.  + i \bar{m} \left( \bar{\psi}_1 \Gamma \psi_1 - \bar{\psi}_2 \Gamma \psi_2 \right)  - i m_B \left( \bar{\psi}_1 \psi_2 + \bar{\psi}_2 \psi_1 \right) \right] \ , 
\end{eqnarray}
with:
\begin{eqnarray*}
ds^2 &=& \frac{R_2^2}{\xi^2} \left( - d \tau^2 + d \xi^2 \right)  \ , \quad \mathcal{D}_\alpha \ = \  \partial_\alpha + \frac{1}{4} \omega_{\alpha \underline{a} \underline{b}} \Gamma^{\underline{a} \underline{b}} - i g A_\alpha \ , \quad A_\tau = \frac{e_d}{\xi} \ , \\
\Gamma^{\underline{\tau}} &=& i \sigma_x  \ , \quad \Gamma^{\underline{\xi}}  =\  - \sigma_z \ , \quad \Gamma \  =  \  - \sigma_y \ , 
\end{eqnarray*}
where we have used the trick of ref.~\cite{Faulkner:2009wj} to write the contribution from the momentum term as a time--reversal violating mass term.  In fact, to recover the action of ref.~\cite{Faulkner:2009wj}, we simply would set $m_B$ to zero.  We emphasize that from our definitions of $m_B$ and $\bar{m}$ in equation \reef{eqt:def} they are always greater than zero.  
Using an ansatz for the fields $\psi_i$ of the form:
\begin{equation}
\psi_{1,2}(\tau, \xi) = e^{- i \omega \tau} \xi^{1/2} \phi_{1,2}(\omega, \xi) \ ,
\end{equation}
the equations of motion reduce to:
\begin{equation} \label{eqt:eom_inner}
  \xi \partial_\xi \phi_{^1_2} + m R_2  \sigma_z \phi_{^1_2} - i \xi \left( \omega + \frac{g e_d}{\xi} \right) \sigma_y \phi_{^1_2} \pm  \bar{m} R_2 \sigma_x \phi_{^1_2} +  i R_2 m_B \sigma_z \phi_{^2_1} = 0 \ .
\end{equation}
This can be turned into exactly the form of equation \reef{eqt:AdS2_v1} by scaling $\xi \to \xi / \omega$.  We can solve these equations analytically, and we refer the reader to appendix \ref{app:solve} for the details.  Here, we simply quote the $\xi \to 0$ limit of the solution when we impose ingoing boundary conditions at $\xi \to \infty$:
\begin{eqnarray} \label{eqt:exact_asymptotic}
\lim_{\xi \to 0} \phi(\xi) &\equiv& \lim_{\xi \to 0} \left( \begin{array}{c} \phi_1 \\ \phi_2 \end{array}\right)  =   A_1 \left( v_1 \xi^{-\nu_-} + v_2 \xi^{\nu_-}  \mathcal{G}_R^{(-)} \right)+ A_2 \left( v_3 \xi^{-\nu_+} +  v_4 \xi^{\nu_+} \mathcal{G}_R^{(+)} \right) \ ,
\end{eqnarray}
where $v_i$ are the eigenvectors (given in equation \reef{eqt:eigenvector_v}) of equation \reef{eqt:eom_inner} in the limit of $\xi \to 0$, and the eigenvalues are written in terms of:
\begin{equation} \label{eqt:nu_pm}
\nu_\pm=  R_2 \left(- \frac{g^2 e_d^2}{ R_2^2}+ m^2 + \bar{m}^2  - m_B^2 \pm 2 i m_B  \sqrt{m^2 + \bar{m}^2} \right)^{1/2}  \ .
\end{equation}
The quantities $\mathcal{G}_R^{(\pm)}$ are given by:
\begin{equation}
\mathcal{G}_R^{(-)} = \left( 2 \omega \right)^{2 \nu_-} e^{- i \pi \nu_-} \frac{ \Gamma(-2 \nu_-) \Gamma( 1 + \nu_- -  i g e_d)}{\Gamma(2 \nu_-) \Gamma(1 - \nu_- - ig e_d)}   \frac{R_2 \left(m + i \bar{m} \right) - i R_2 m_B P + \nu_- + i  g e_d }{ R_2 \left(m + i \bar{m} \right) - i R_2 m_B P - \nu_- + i  g e_d } \ ,
\end{equation}
\begin{equation}
\mathcal{G}_R^{(+)} =   \left( 2 \omega \right)^{2 \nu_+} e^{- i \pi \nu_+} \frac{ \Gamma(-2 \nu_+) \Gamma( 1 + \nu_+ -  i g e_d)}{\Gamma(2 \nu_+) \Gamma(1 - \nu_+ - ig e_d)}  \frac{R_2 \left(m + i \bar{m} \right) + i R_2 m_B P + \nu_+ + i  g e_d }{ R_2 \left(m + i \bar{m} \right) + i R_2 m_B P - \nu_+ + i  g e_d } \ .
\end{equation}
If we impose outgoing wave boundary conditions at $\xi \to \infty$, the asymptotic solution has the form of equation \reef{eqt:exact_asymptotic}, but with $\mathcal{G}_R^{(\pm)}$ replaced by $\mathcal{G}_A^{(\pm)}$:
\begin{equation}
\mathcal{G}_A^{(-)} = \left( 2 \omega \right)^{2 \nu_-} e^{ i \pi \nu_-} \frac{ \Gamma(-2 \nu_-) \Gamma( 1 + \nu_- +  i g e_d)}{\Gamma(2 \nu_-) \Gamma(1 - \nu_- + ig e_d)}  \frac{R_2 \left(m - i \bar{m} \right) - i R_2 m_B \bar{P} + \nu_- - i  g e_d }{ R_2 \left(m - i \bar{m} \right) - i R_2 m_B \bar{P} - \nu_- - i  g e_d } \ ,
\end{equation}
\begin{equation}
\mathcal{G}_A^{(+)} =   \left( 2 \omega \right)^{2 \nu_+} e^{i \pi \nu_+} \frac{ \Gamma(-2 \nu_+) \Gamma( 1 + \nu_+ +  i g e_d)}{\Gamma(2 \nu_+) \Gamma(1 - \nu_+ +ig e_d)}  \frac{R_2 \left(m - i \bar{m} \right) + i R_2 m_B \bar{P} + \nu_+ - i  g e_d }{ R_2 \left(m - i \bar{m} \right) + i R_2 m_B \bar{P} - \nu_+ - i  g e_d } \ .
\end{equation}
%
\section{Properties of the Green's Function} \label{sec:speculate}
%
At this point, we assume that we can extract the source and vev of the dual operator from the bulk fermion.  
This is not a trivial assumption since the asymptotic behavior of the fields is very different in these gravity duals of non--commutative Yang--Mills theory.  For example, the asymptotic behavior of the scalar field is shown in ref.~\cite{Maldacena:1999mh} to be in terms of exponentials and not powers of the AdS radial coordinate.
With this assumption in mind, we follow the analysis of ref.~\cite{Faulkner:2009wj} as to how to proceed.  We separate the AdS radial direction into an ``inner'' and ``outer'' region, with fermionic fields $\phi_I$ and $\phi_O$ in each respectively.  The equation of motion in the outer region is given by equation \reef{eqt:eom_outer} and that in the inner region is given by \reef{eqt:eom_inner}.  If we take $\omega = 0$ in the outer region and study the limit of $z \to z_0$, we find that it matches the limit of $\xi \to 0$ of the equation of motion in the inner region.  Therefore, at the intersection of the two regions, we make the following identification:
\begin{equation}
\phi_O( \omega = 0, z \to z_0) = \phi_I ( \omega, \xi \to 0) \ .
\end{equation}
Furthermore, by assuming the following form for $\phi_O$:
\begin{equation}
\phi_O(\omega, z) = \eta_1(\omega,z) + \mathcal{G}_R^{(-)}(\omega) \eta_2 (\omega,z) + \eta_3(\omega,z) + \mathcal{G}_R^{(+)}(\omega) \eta_4(\omega,z) \ ,
\end{equation}
and expanding $\eta_i$ in powers of $\omega$ such as:
\begin{equation}
\eta_i(\omega, z) = \eta_i^{(0)}(z) + \omega \eta_i^{(1)} (z) + \dots \ , 
\end{equation}
we can write the boundary condition at the intersection of the inner and outer region as follows:
\begin{equation*}
\eta_1^{(0)}(z \to z_0) = v_1 \left( \frac{\sqrt{3}}{12} \frac{1}{z_0 - z} \right)^{- \nu_-} \ , \quad \eta_2^{(0)}( z \to z_0) = v_2 \left( \frac{\sqrt{3}}{12} \frac{1}{z_0 - z} \right)^{\nu_-} \ , 
\end{equation*}
\begin{equation} \label{eqt:bc}
\eta_3^{(0)}( z \to z_0) = v_3 \left( \frac{\sqrt{3}}{12} \frac{1}{z_0 - z} \right)^{- \nu_+} \ , \quad \eta_4^{(0)}(z \to z_0) = v_4 \left( \frac{\sqrt{3}}{12} \frac{1}{z_0 - z} \right)^{\nu_+} \ .
\end{equation}
We can find the boundary conditions for the higher order terms $\eta_i^{(n)}$ for $n>0$ by solving the equation of motion in the outer region perturbatively in $\omega$.  Since each $\eta_i^{(0)}$ corresponds to an independent boundary condition, we can solve the equation of motion in the outer region for each one separately.  Therefore, one can presumably write a similar expression for the retarded Green's function (assuming the prescription for sources and vevs remains the same) as in ref.~\cite{Faulkner:2009wj} as:
\begin{eqnarray}
G_R^{(-)} &=& K \frac{ b_1^{(0)} + b_1^{(1)} \omega + O(\omega^2) + \mathcal{G}_R^{(-)}(\omega) \left( b_2^{(0)} + b_2^{(1)} \omega + O(\omega^2) \right)}{a_1^{(0)} + a_1^{(1)} \omega + O(\omega^2) + \mathcal{G}_R^{(-)}(\omega) \left( a_2^{(0)} + a_2^{(1)} \omega + O(\omega^2) \right)} \ , \\
G_R^{(+)} &=& K \frac{ b_3^{(0)} + b_3^{(1)} \omega + O(\omega^2) + \mathcal{G}_R^{(+)}(\omega) \left( b_4^{(0)} + b_4^{(1)} \omega + O(\omega^2) \right)}{a_3^{(0)} + a_3^{(1)} \omega + O(\omega^2) + \mathcal{G}_R^{(+)}(\omega) \left( a_4^{(0)} + a_4^{(1)} \omega + O(\omega^2) \right)} \ ,
\end{eqnarray}
where $K$ is a constant and $(a_i^{(n)}, b_i^{(n)})$'s are proportional to the sources and vevs extracted from $\eta_i^{(n)}$ at $z \to 0$.   For small $\omega$, we write:
\begin{equation}
G_R^{(-)} \approx K \frac{ b_1^{(0)}  + \mathcal{G}_R^{(-)}(\omega) b_2^{(0)} }{a_1^{(0)} + \mathcal{G}_R^{(-)}(\omega) a_2^{(0)} } \ , \quad G_R^{(+)} \approx K \frac{ b_3^{(0)} + \mathcal{G}_R^{(+)}(\omega) b_4^{(0)} }{a_3^{(0)}  + \mathcal{G}_R^{(+)}(\omega)a_4^{(0)} } \ .
\end{equation}
If we define:
\begin{equation}
| R_2^2 \left( m^2 + \bar{m}^2 - m_B^2 \right) - g^2 e_d^2 | \pm 2 i R_2^2 m_B \sqrt{m^2 + \bar{m}^2} = \nu  e^{i \theta_\pm} \ ,
\end{equation}
we can write:
\begin{equation}
\nu_\pm = \left\{\begin{array}{cl}
-i \nu^{1/2} e^{ i \theta_\mp /2} \ , & \mathrm{when} \ R_2^2 \left( m^2 + \bar{m}^2 - m_B^2 \right) - g^2 e_d^2 < 0 \\
 \nu^{1/2} e^{i \theta_\pm /2} \ , & \mathrm{when} \ R_2^2 \left( m^2 + \bar{m}^2 - m_B^2 \right) - g^2 e_d^2 > 0
 \end{array} \right.  \ .
\end{equation}
The phase $\theta_+$ ranges from $0$ to $\pi/2$, and $\theta_-$ ranges from $- \pi/2$ to $0$.  When $m_B = 0$, we have that $\theta_\pm = 0$, and we recover the results of ref.~\cite{Faulkner:2009wj} where $\nu_\pm$ are either purely imaginary (in the oscillatory region) or purely real (in the non--oscillatory region).  
%
%
\subsection{Oscillatory Region}
%
For the case of  $R_2^2 \left( m^2 + \bar{m}^2 - m_B^2 \right) - g^2 e_d^2 < 0$, the complex $\nu_\pm$ exist in the third and fourth quadrant of the complex plane respectively.  This region of parameter space is the generalization of the oscillatory region of ref.~\cite{Faulkner:2009wj}.  The complex $\nu_\pm$ and the functions $\mathcal{G}_R^{(\pm)}$ satisfy the following properties:
\begin{equation}
\nu_\pm^\ast = - \nu_\mp \ , \quad \left(\mathcal{G}_R^{(\pm)}(\omega)\right)^\ast = \left( \mathcal{G}_A^{(\mp)} (\omega^{\ast}) \right)^{-1} \ .
\end{equation}
To make some headway, let us consider the case of $m = 0$.  For this choice, we have:
\begin{equation}
v_{^1_4}^{\ast} = \left( \begin{array}{cc}  
-1_{2 \times 2} & 0 \\
0 & 1_{2 \times 2} 
\end{array} \right) v_{^4_1} \ , \quad v_{^2_3}^{\ast} = \left( \begin{array}{cc}  
-1_{2 \times 2} & 0 \\
0 & 1_{2 \times 2} 
\end{array} \right) v_{^3_2} \ .
\end{equation}
This property under complex conjugation of the eigenvectors is similar to that of equation \reef{eqt:cc}, which we found restores the complex conjugated equations of motion to their original form.  Using this result, if we complex  conjugate the initial conditions in equation \reef{eqt:bc}, we find the following relationship between the fields in the outer region:
\begin{equation}
\eta_{^1_4}^\ast \to \eta_{^4_1} \ , \quad \eta_{^2_3}^\ast \to \eta_{^3_2} \ .
\end{equation}
We conclude that we must have the following relationship between the sources and vevs:
\begin{equation} \label{eqt:oscillatory}
a_{^1_4}^\ast = a_{^4_1} \ , \quad a_{^2_3}^\ast = a_{^3_2} \ , \quad b_{^1_4} ^\ast = b_{^4_1} \ , \quad b_{^2_3}^\ast = b_{^3_2} \ ,
\end{equation}
Using these results, we write the retarded Green's function as follows:
\begin{equation}
G_R^{(-)} (\omega) \approx K \frac{b_1^{(0)}}{a_1^{(0)}} \frac{ 1 + \left( \omega^{2 \nu_-} c_R^{(-)}  \frac{b_2^{(0)}}{b_1^{(0)}} + \omega^{-2 \nu_+} (c_A^{(+)})^{-1}  \frac{a_3^{(0)}}{a_4^{(0)}} \right) + \omega^{4 \mathrm{Re}( \nu_-)} \frac{b_2^{(0)} a_3^{(0)}}{b_1^{(0)} a_4^{(0)}} (c_A^{(+)})^{-1} c_R^{(-)}  }{ | 1 +  \omega^{2 \nu_-} c_R^{(-)}  \frac{a_2^{(0)}}{a_1^{(0)}} |^2 } \ , 
\end{equation}
\begin{equation} \label{eqt:G_R+}
G_R^{(+)} (\omega) \approx K \frac{b_4^{(0)}}{a_4^{(0)}} \frac{ 1 + \left( \omega^{-2 \nu_+} (c_R^{(+)})^{-1}  \frac{b_3^{(0)}}{b_4^{(0)}} + \omega^{2 \nu_-} c_A^{(-)}  \frac{a_2^{(0)}}{a_1^{(0)}} \right) + \omega^{-4 \mathrm{Re}( \nu_+)} \frac{b_3^{(0)} a_2^{(0)}}{b_4^{(0)} a_1^{(0)}} (c_R^{(+)})^{-1} c_A^{(-)}  }{ | 1 + \omega^{- 2 \nu_+} (c_R^{(+)})^{-1}  \frac{a_3^{(0)}}{a_4^{(0)}} |^2 } \ ,
\end{equation}
where we have defined:
\begin{equation}
\mathcal{G}_R^{(\pm)} = \omega^{2\nu_\pm} c_R^{(\pm)}  \ , \quad \mathcal{G}_A^{(\pm)} =  \omega^{2 \nu_\pm} c_A^{(\pm)} \ .
\end{equation}
and $(c_R^{(\pm)}, c_A^{(\pm)})$ do not depend on $\omega$.  In calculating the expression for $G_R^{(+)}$ in equation \reef{eqt:G_R+}, we have used the fact that $\mathcal{G}_R^{(+)}$ diverges for small $\omega$ because $\nu_+$ has a negative real part.  Setting $\omega$ to zero, we find:
\begin{equation}
G_R^{(-)} (\omega = 0) = K \frac{b_1}{a_1} \ , \quad G_R^{(+)} (\omega = 0) = K \frac{b_4}{a_4} \ ,
\end{equation}
and using equation \reef{eqt:oscillatory}, we have our main result that:
\begin{equation} \label{eqt:negative_imag1}
\left( G_R^{(-)} (\omega = 0) \right)^\ast = G_R^{(+)} (\omega  = 0) \ .
\end{equation}
The key reason for this result is that when $m_B \neq 0$, $\mathrm{Re}( \nu_+) < 0$ so $\mathcal{G}_R^{(+)}( \omega \to 0)$ diverges.  This behavior is absent in the case of $m_B = 0$, where the conclusion of equation \reef{eqt:negative_imag1} is not reached. 
Since the equation of motion \reef{eqt:eom_outer} is complex, we can imagine that generically, the coefficients $(a_i, b_i)$ are complex, and we can expect that one of the $G_R^{(\pm)}$ has negative imaginary part by equation \reef{eqt:negative_imag1}.  A negative spectral function is prohibited by unitarity, so we are led to the conclusion that  in the oscillatory regime when $m_B \neq 0$, the limit of $\omega = 0$ is not allowed, \emph{i.e.} to study the case of $\mathrm{Re}(\omega) \to 0$ in the gravity dual, one should have a finite imaginary part for $\omega$.  
%
\subsection{Non--Oscillatory Region}
%
For the case when $R_2^2 \left( m^2 + \bar{m}^2 - m_B^2 \right) - g^2 e_d^2 \geq 0$, the complex $\nu_\pm$ exist in the first and fourth quadrant of the complex plane respectively.  This case is a generalization of the non--oscillatory region of ref.~\cite{Faulkner:2009wj}.  The complex $\nu_\pm$ and $\mathcal{G}_R^{(\pm)}$ satisfy the following property:
\begin{equation}
\nu_\pm^\ast =  \nu_\mp  \ , \quad \left(\mathcal{G}_R^{(\pm)}(\omega)\right)^\ast = \left( \mathcal{G}_A^{(\mp)} (\omega^{\ast}) \right) \ .
\end{equation}
We consider again the case of $m = 0$.  For this choice, we have:
\begin{equation} \label{eqt:complex_vi}
v_{^1_3}^{\ast} = \left( \begin{array}{cc}  
-1_{2 \times 2} & 0 \\
0 & 1_{2 \times 2} 
\end{array} \right) v_{^3_1} \ , \quad v_{^2_4}^{\ast} = \left( \begin{array}{cc}  
-1_{2 \times 2} & 0 \\
0 & 1_{2 \times 2} 
\end{array} \right) v_{^4_2} \ .
\end{equation}
Following the same arguments from the previous section we have that:
\begin{equation}
a_{^1_3}^\ast = a_{^3_1} \ , \quad a_{^2_4}^\ast = a_{^4_2} \ , \quad b_{^1_3}^\ast = b_{^3_1} \ , \quad b_{^2_4}^\ast = b_{^4_2} \ .
\end{equation}
Calculating the retarded Green's function gives:
\begin{equation}
G_R^{(-)} (\omega) \approx K \frac{b_1^{(0)}}{a_1^{(0)}} \frac{ 1 + \left( \omega^{2 \nu_-} c_R^{(-)}  \frac{b_2^{(0)}}{b_1^{(0)}} + \omega^{2 \nu_+} c_A^{(+)}  \frac{a_4^{(0)}}{a_3^{(0)}} \right) + \omega^{4 \mathrm{Re}( \nu_-)} \frac{b_2^{(0)} a_4^{(0)}}{b_1^{(0)} a_3^{(0)}} c_A^{(+)} c_R^{(-)}  }{ | 1 + \omega^{2 \nu_-} c_R^{(-)}  \frac{a_2^{(0)}}{a_1^{(0)}} |^2 } \ ,
\end{equation}
\begin{equation}
G_R^{(-)} (\omega) \approx K \frac{b_3^{(0)}}{a_3^{(0)}} \frac{ 1 + \left( \omega^{2 \nu_+} c_R^{(+)}  \frac{b_4^{(0)}}{b_3^{(0)}} + \omega^{2 \nu_-} c_A^{(-)}  \frac{a_2^{(0)}}{a_1^{(0)}} \right) + \omega^{4 \mathrm{Re}( \nu_+)} \frac{b_4^{(0)} a_2^{(0)}}{b_3^{(0)} a_1^{(0)}} c_A^{(-)} c_R^{(+)}  }{ | 1 +  \omega^{2 \nu_+} c_R^{(+)}  \frac{a_4^{(0)}}{a_3^{(0)}} |^2 } \ .
\end{equation}
Again, we find that :
\begin{equation} \label{eqt:non-oscillatory}
\left( G_R^{(-)}(\omega = 0) \right)^{\ast}= G_R^{(+)}(\omega = 0) \ .
\end{equation}
Recall again that  when $m_B \neq 0$, the equation of motion is complex, so generically $(a_i, b_i)$ are complex and $\mathrm{Im}( G^{(\pm)}_R (\omega = 0)) $ have opposite signs (if the terms are non--zero).  Therefore, we are led to the same conclusion as that of the previous section that, by unitarity, we must have a finite imaginary part for $\omega$ if the real part is zero.  We reiterate that when $m_B = 0$, the boundary conditions and equations of motion are real so the coefficients $(a_i, b_i)$ are real as well; then equation \reef{eqt:non-oscillatory} is simply the statement that the imaginary part of the retarded Green's function is zero.
%
\section{Conclusion} \label{sec:conclusions}
%
We have studied a probe fermion charged under a $U(1)$ gauge field and a two form potential $B_{(2)}$ in a gravity background dual to strongly coupled non--commutative Yang--Mills theory in the 't Hooft limit with $U(1)$ global currents turned on.  The background is a five dimensional solution of an effective action we construct, where the five dimensional solution is motivated by a new ten dimensional solution of type IIB supergravity that we found.
In the near horizon AdS$_2$ region, the $B_{(2)}$ potential appears as an interaction term between two fermions (which form the full spinor in the AdS$_5$ space).  The two operator dimensions $\nu_\pm$ extracted from the asymptotic behavior in the AdS$_2$ region are complex (as opposed to being simply pure real or pure imaginary), and they lie either in the third and fourth quadrant (in the oscillatory region) or the first and fourth quadrant  (in the non--oscillatory region) of the complex plane.  By assuming that the Green's function in the AdS$_5$ space retains the same form as that from the usual AdS/CFT dictionary and using properties of the equations of motion, we find that the two elements $G_R^{(\pm)}$ of the retarded Green's function are complex conjugates of each other at $\omega = 0$ in both the oscillatory and non--oscillatory regions.  Therefore, one expects generically that  one of the elements of the retarded Green's function has negative imaginary part at $\omega = 0$.  This violates unitarity, and leads us to the conclusion that in order to study the zero frequency limit in the dual gauge theory, one has to have a finite imaginary part for $\omega$ as the real part is sent to zero in the dual gravity description.  In other words, at $\mathrm{Re}(\omega) = 0$, if a quasinormal mode exists, it must have a finite lifetime (if $\mathrm{Im}(\omega_\ast) < 0$) or be unstable (if $\mathrm{Im}(\omega_\ast) > 0$).
A novel property of our setup is that it is an example of a case where the dynamics in the transverse space to the AdS$_2$ can greatly modify the behavior of the retarded Green's function in the $\omega \to 0$ limit.   The flux in the $(x_2,x_3)$ directions breaks the $SO(3)$ symmetry of the three spatial directions in the dual Yang--Mills theory, which suggests that the underlying Fermi surface (if it exists) is not spherical in shape.  Our ansatz only explores the $k_1 = 0$ slice of this surface, but a more general ansatz could explore the full shape of the surface.  Furthermore, the dynamics in the transverse space give rise to the result that excitations with infinite lifetimes must occur at finite $\omega$.  If we identify $\omega$ as the energy above/below the chemical potential, this result implies that the Fermi energy is not equal to the chemical potential.  Modes with this behavior were found in the work of refs.~\cite{Albash:2009wz,Albash:2010yr}, where the fermion was coupled to a $U(1)$ magnetic field.  Our study provides a framework to perhaps understand why zero frequency modes did not appear in that work.  The presence of the magnetic field may have forced the dimension of the operators in the AdS$_2$ region to be complex, as the system in this paper does.  This would push the long lived excitations to have energies above the chemical potential rather than at the chemical potential. 
\section*{Acknowledgements}
We would like to thank Nikolay Bobev for numerous useful discussions in addition for comments to the manuscript.  We would also like to thank Clifford Johnson for comments to the manuscript.  This work was supported by the US Department of Energy.

\appendix
\section{Solving the Equations of Motion} \label{app:solve}
%
The asymptotic $\xi \to 0$ behavior of the equations of motion given in equation \reef{eqt:eom_inner} is:
\begin{eqnarray}
\xi \partial_{\xi} \phi_1 &=& - R_2  \left( \begin{array}{cc} m   & \bar{m} - \frac{g e_d}{R_2} \\  \bar{m} + \frac{g e_d}{R_2} &  - m  \end{array} \right) \phi_1 - R_2  \left( \begin{array}{cc} i m_B  & 0 \\ 0&  -i m_B \end{array} \right) \phi_2 \ , \nonumber  \\
\xi \partial_{\xi} \phi_2 &=& - R_2  \left( \begin{array}{cc} m   & -\bar{m} - \frac{g e_d}{R_2 } \\  -\bar{m} + \frac{g e_d}{R_2} &  - m \end{array} \right) \phi_2 - R_2  \left( \begin{array}{cc} i m_B  & 0 \\ 0 &  -i m_B \end{array} \right) \phi_1 \ .
\end{eqnarray}
Solving these equations reduces to finding the eigenvalues and eigenvectors of a $4 \times 4$ matrix.  The solution is given by:
\begin{equation}
\phi \equiv \left( \begin{array}{c} \phi_1 \\ \phi_2 \end{array} \right) = A_1 \xi^{- \nu_-} v_1 + A_2 \xi^{\nu_-} v_2 +A_3\xi^{-\nu_+} v_3 + A_4 \xi^{\nu_+} v_4 \ , 
\end{equation}
where $v_i$ are the eigenvectors given by:
\begin{equation*}
v_1 = \left( \begin{array}{c}
- \beta_+^{(-)} \\
- \alpha_-  \\
\gamma_+^{(-)}\\
m g e_d - \nu_- \bar{m}
\end{array} \right) \ , \quad
v_2 = \left( \begin{array}{c}
 -  \beta_-^{(-)} \\
 -  \alpha_- \\
\gamma_-^{(-)}\\
m g e_d + \nu_- \bar{m}
\end{array} \right) \ , 
\end{equation*}
\begin{equation} \label{eqt:eigenvector_v}
v_3 = \left( \begin{array}{c} 
 \beta_+^{(+)} \\
  \alpha_+ \\
\gamma_+^{(+)}\\
m g e_d - \nu_+ \bar{m}
  \end{array} \right) \ , \quad
v_4 = \left( \begin{array}{c} 
\beta_-^{(+)} \\
 \alpha_+  \\
\gamma_-^{(+)}\\
m g e_d + \nu_+ \bar{m}
\end{array} \right) \ , 
\end{equation}
and we have defined:
\begin{eqnarray}
\nu_\pm&=&  R_2 \left(- \frac{g^2 e_d^2}{ R_2^2}+ m^2 + \bar{m}^2  - m_B^2 \pm 2 i m_B  \sqrt{m^2 + \bar{m}^2} \right)^{1/2}   \ , \nonumber \\
\alpha_\pm &=&   \left( \sqrt{ m^2 + \bar{m}^2} \left( g e_d + R \bar{m} \right) \pm i R \bar{m} m_B \right)  \ , \nonumber \\
\beta_\pm^{(+)} &=&  \left( \sqrt{ m^2 + \bar{m}^2} \left( R m \pm \nu_+ \right) + i R m m_B \right)  \ , \nonumber  \\
\beta_\pm^{(-)} &=&  \left( \sqrt{ m^2 + \bar{m}^2} \left( R m \pm \nu_- \right) - i R m m_B \right)  \ , \nonumber  \\
\gamma_\pm^{(+)} &=& R(m^2 + \bar{m}^2) + g e_d \bar{m} + i R m_B \sqrt{m^2 + \bar{m}^2} \pm m \nu_+ \ , \nonumber \\
\gamma_\pm^{(-)} &=& R(m^2 + \bar{m}^2) + g e_d \bar{m} - i R m_B \sqrt{m^2 + \bar{m}^2} \pm m \nu_- \ . \nonumber
\end{eqnarray}
The eigenvectors $(v_1,v_2)$ are linearly independent from $(v_3,v_4)$; however, $v_1$ is not linearly independent from $v_2$, and $v_3$ is not linearly independent from $v_4$.   When $m_B = 0$, the four eigenvalues reduce to two distinct eigenvalues with multiplicity two, and our choice of eigenvectors is degenerate and must be chosen differently.   We emphasize that we choose a different basis for our eigenvectors than ref.~\cite{Faulkner:2009wj}.
It is convenient to write the equations of motion in terms of a second order differential equation given by:
\begin{eqnarray}
\xi^2 \partial_\xi^2 \phi + \xi \partial_\xi \phi  &=& \left( R_2^2 \left( m^2 + \bar{m}^2 - m_B^2 \right) - \left( \xi \omega + g e_d \right)^2 \right)  \left( \begin{array} {cc}1 & 0 \\ 0 & 1 \end{array} \right) \phi \nonumber \\
&&+ \left( \begin{array}{cc}  i \omega \xi \sigma_y & - 2  R^2_2 m m_B - 2 i R_2^2 m \bar{m} \sigma_y   \\ -2  R^2_2 m m_B  + 2 i R_2^2 m \bar{m} \sigma_y &  i \omega \xi \sigma_y  \end{array} \right) \phi \ .
\end{eqnarray}
To diagonalize the equations, we change our basis as follows:
\begin{equation}
\phi (\xi) = \mathcal{P} \tilde{\phi} (\xi) \ , 
\end{equation}
with the matrix $\mathcal{P}$ given by:
\begin{equation}
\mathcal{P} = \left(\begin{array}{cccc} 
 -i P & i \bar{P} & i  & - i \\ 
-P & - \bar{P} & 1  & 1 \\
i &  -i & i P^{-1} & -i \bar{P}^{-1} \\
1 & 1 & P^{-1} & \bar{P}^{-1}
\end{array} \right) \ ,
\end{equation}
where:
\begin{equation} 
P = \sqrt{ \frac{ m + i \bar{m}}{ m - i \bar{m}}} \ , \quad \bar{P} = \sqrt{ \frac{ m - i \bar{m}}{ m + i \bar{m}}} \ .
\end{equation}
This leads to the following diagonalized equations of motion:
\begin{eqnarray}
\xi^2 \partial_\xi^2 \tilde{\phi}_1  + \xi \partial_\xi \tilde{\phi}_1 &=&   \left( R_2^2 \left( \sqrt{m^2 + \bar{m}^2} - i m_B \right)^2 - i \omega \xi-  \left( \xi \omega + g e_d \right)^2 \right) \tilde{ \phi}_1 \ , \nonumber \\
\xi^2 \partial_\xi^2 \tilde{\phi}_2  + \xi \partial_\xi \tilde{\phi}_2 &=&   \left( R_2^2 \left( \sqrt{m^2 + \bar{m}^2} - i m_B \right)^2 + i \omega \xi -  \left( \xi \omega + g e_d \right)^2 \right) \tilde{ \phi}_2 \ , \nonumber \\
\xi^2 \partial_\xi^2 \tilde{\phi}_3  + \xi \partial_\xi \tilde{\phi}_3 &=&   \left( R_2^2 \left( \sqrt{m^2 + \bar{m}^2} + i m_B \right)^2 - i \omega \xi -  \left( \xi \omega + g e_d \right)^2 \right) \tilde{ \phi}_3 \ , \nonumber \\
\xi^2 \partial_\xi^2 \tilde{\phi}_4  + \xi \partial_\xi \tilde{\phi}_4 &=&   \left( R_2^2 \left( \sqrt{m^2 + \bar{m}^2} + i m_B \right)^2 + i \omega \xi -  \left( \xi \omega + g e_d \right)^2 \right) \tilde{ \phi}_4 \ .
\end{eqnarray}
Let us make the substitution:
\begin{equation}
\tilde{\phi}_1 = e^{i \omega \xi} \xi^{- \nu_-} \tilde{\psi}_1 \ , \quad \tilde{\phi}_2 = e^{i \omega \xi} \xi^{- \nu_-} \tilde{\psi}_2 \ , \quad \tilde{\phi}_3 = e^{i \omega \xi} \xi^{- \nu_+} \tilde{\psi}_3 \ , \quad  \tilde{\phi}_4 = e^{i \omega \xi} \xi^{- \nu_+} \tilde{\psi}_4 \ .
\end{equation}
We are choosing ingoing wave boundary conditions at $\xi \to \infty$ since we are interested in calculating the retarded Green's function \cite{Son:2002sd,Iqbal:2009fd}.  With this substitution, the equations of motion are now given by:
\begin{eqnarray}
& \xi \partial_\xi^2 \tilde{\psi}_1 + \left( 1 - 2 \nu_- + 2 i \omega \xi \right) \partial_\xi \tilde{\psi}_1 + 2 i \omega \left( 1 - i g e_d - \nu_- \right) \tilde{\psi}_1 = 0 \ , & \nonumber \\
& \xi \partial_\xi^2 \tilde{\psi}_2 + \left( 1 - 2 \nu_- + 2 i \omega \xi \right) \partial_\xi \tilde{\psi}_2 + 2 i \omega \left( - i g e_d - \nu_- \right) \tilde{\psi}_2 = 0 \ , &  \nonumber \\
& \xi \partial_\xi^2 \tilde{\psi}_3 + \left( 1 - 2 \nu_+ + 2 i \omega \xi \right) \partial_\xi \tilde{\psi}_3 + 2 i \omega \left( 1 - i g e_d - \nu_+ \right) \tilde{\psi}_3 = 0 \ , & \nonumber \\
& \xi \partial_\xi^2 \tilde{\psi}_4+ \left( 1 - 2 \nu_+ + 2 i \omega \xi \right) \partial_\xi \tilde{\psi}_4 + 2 i \omega \left( - i g e_d - \nu_+ \right) \tilde{\psi}_4 = 0 \ .  &
\end{eqnarray}
Changing variables to $\zeta = - 2 i \omega \xi$, we recover Kummer's equations for all four fields:
\begin{equation} \label{eqt:Kummer}
\zeta \partial_\zeta^2 \tilde{\psi}_i + \left( b_i - \zeta \right) \partial_\zeta \tilde{\psi}_i - a_i \tilde{\psi}_i = 0 \ ,
\end{equation}
with:
\begin{equation}
a_1 = 1 - i g e_d - \nu_- \ , \quad a_2 = - i g e_d - \nu_- \ , \quad a_3 = 1 - i g e_d - \nu_+ \ , \quad  a_4 = - i g e_d - \nu_+ \ , \nonumber
\end{equation}
\begin{equation}
b_1 = b_2 = 1 - 2 \nu_- \ , \quad b_3 = b_4 = 1 - 2 \nu_+ \ . \nonumber
\end{equation}
The solutions to equation \reef{eqt:Kummer} are given by:
\begin{equation}
\tilde{\psi}_i = c_i \left( M(a_i, b_i, \zeta) + d_i \zeta^{1- b_i} M( a_i - b_i + 1, 2 - b_i, \zeta) \right) \ ,
\end{equation}
where $M(a,b,z)$ is Kummer's function of the first kind, also known as the confluent hypergeometric function\footnote{For properties of the confluent hypergeometric function, see for example ref.~\cite{Arfken}}.  Its normalization is such that $M(a,b,0) = 1$.  To match to the eigenvectors given in equation \reef{eqt:eigenvector_v}, we take:
\begin{eqnarray*}
c_1 &=& - i \frac{A_1 \left( m - i \bar{m} \right) }{2} \left( R \left( m + i \bar{m} \right) + i g e_d - i R m_B P + \nu_- \right) \ , \\
c_2 &=&  i \frac{A_1 \left( m + i \bar{m} \right)}{2} \left( R \left( m - i \bar{m} \right) - i g e_d - i R m_B \bar{P} + \nu_- \right) \ , \\
c_3 &=& - i \frac{A_2 \sqrt{m^2 + \bar{m}^2}}{2} \left( R \left( m + i \bar{m} \right) + i g e_d + i R m_B P + \nu_+ \right) \ , \\
c_4 &=&  i \frac{A_2\sqrt{m^2 + \bar{m}^2}}{2 } \left( R \left( m - i \bar{m} \right) - i g e_d + i R m_B \bar{P} + \nu_+ \right) \ , \\
\end{eqnarray*}
and for regularity of the solution at $\xi \to \infty$, we take:
\begin{equation*}
d_i = \frac{ \Gamma(b_i - 1) \Gamma(1 + a_i - b_i)}{\Gamma(1 - b_i) \Gamma(a_i)} \ .
\end{equation*}
With these choices, we find that the solution has $\xi \to 0$ asymptotic behavior given by equation \reef{eqt:exact_asymptotic}.  We can calculate the advanced Green's function in a similar fashion.  Instead of the ingoing wave boundary conditions, we choose outgoing wave boundary conditions for our fields, \emph{i.e.}:
\begin{equation}
\tilde{\phi}_1 = e^{-i \omega \xi} \xi^{- \nu_-} \tilde{\psi}_1 \ , \quad \tilde{\phi}_2 = e^{-i \omega \xi} \xi^{- \nu_+} \tilde{\psi}_2 \ , \quad \tilde{\phi}_3 = e^{-i \omega \xi} \xi^{- \nu_-} \tilde{\psi}_3 \ , \quad  \tilde{\phi}_4 = e^{-i \omega \xi} \xi^{- \nu_+} \tilde{\psi}_4 \ .
\end{equation}
\providecommand{\href}[2]{#2}\begingroup\raggedright\endgroup
\end{document}